\documentclass[12pt]{article}
\usepackage{amssymb}

\usepackage{natbib,graphicx,setspace,lscape,longtable}
\usepackage{epsfig}
\usepackage{amsmath,amsthm,bm,color}

\usepackage{xcolor}         
\usepackage{amsmath}                        
\usepackage{graphicx}               
\usepackage{caption}
\usepackage{subcaption}
\usepackage{hyperref}
\usepackage{multirow}

\definecolor{DarkGreen}{rgb}{0.5,0.8,0.6}   
\definecolor{RGBblack}{rgb}{0.0,0.0,0.0}    

\newcommand{\hei}[1]{\color{black}{#1} \color{black} }
\definecolor{grau}{rgb}{0.8,0.8,0.8}

\newcommand{\chen}[1]{\color{orange}}

\def\hei{\color{black}}

\newcommand{\jj}{\hei \rm}

\newcommand{\ech}{\hei\rm}

\newcommand{\is}{\itemsep=0pt}
\newcommand{\bd}[1]{\begin{description}[#1]\is} 
  \newcommand{\ed}{\end{description}}
\newcommand{\bi}{\begin{itemize}\is}
  \newcommand{\ei}{\end{itemize}}
\newcommand{\be}{\begin{enumerate}\is}
  \newcommand{\ee}{\end{enumerate}}

\makeatletter
\newcommand*{\rom}[1]{\expandafter\@slowromancap\romannumeral #1@}

\newcommand{\bX}{\bm X}
\newcommand{\bx}{\bm x}
\newcommand{\bPi}{\Pi}
\newcommand{\cPi}{\bm \Pi}
\newcommand{\by}{\bm y}

\newcommand{\bz}{\bm z}
\newcommand{\btheta}{\bm \theta}

\newcommand{\pPir}{p(\Pi_r)}  
\newcommand{\Pir}{\Pi_r}  
\renewcommand{\Pr}{p}
\newcommand{\zs}{z^\star}

\renewcommand{\Re}{\mathbb{R}}
\renewcommand{\th}{\theta}
\newcommand{\bth}{\btheta}

\newcommand{\byn}{\by^{(n)}}
\newcommand{\bXn}{\bX^{(n)}}
\newcommand{\bzn}{\bz^{(n)}}
\newcommand{\Be}{\mbox{Be}}
\newcommand{\Unif}{\mbox{Unif}}
\newcommand{\xno}{\bx_{n+1}}
\newcommand{\Dn}{\mathcal{D}_n}
\newcommand{\tbx}{\tilde{\bx}}
\newcommand{\tx}{\tilde{x}}
\newcommand{\Sn}{S^0}

\begin{document}
\begin{center}
{\bf\Large Subgroup-Based Adaptive (SUBA) Designs for Multi-Arm Biomarker Trials}\\
\bigskip
\bigskip

{\it\large Yanxun Xu$^{1}$, Lorenzo Trippa$^{2}$, Peter M\"uller$^{3}$and Yuan Ji$\;^{4,5}$}

\bigskip
\end{center}

\vskip 2in

\centerline{1 Division of Statistics and Scientific Computing, 
The University of Texas at Austin, Austin, TX, U.S.A.}
\centerline{2 Department of Biostatistics, Harvard School of Public Health, Boston, MA, U.S.A.}
\centerline{3 Department of Mathematics, The University of Texas at Austin, Austin, TX, U.S.A.}
\centerline{4 Center for Clinical and Research Informatics, NorthShore University HealthSystem Evanston, IL, U.S.A}
\centerline{5 Prytzker School of Medicine, The University of Chicago, Chicago, IL, U.S.A}

\vskip 3in 

Email: yji@health.bsd.uchicago.edu

\clearpage
\newpage

\begin{abstract}
Targeted therapies based on biomarker profiling are becoming a
mainstream direction of cancer research and treatment. Depending on
the expression of specific prognostic biomarkers, 
targeted therapies assign 
different cancer drugs to subgroups of
patients even if they are diagnosed with the same type of cancer by
traditional means, such as tumor location. For example, Herceptin is
only indicated for 
the subgroup of patients with HER2+ breast cancer,
but not other types of breast cancer.  However, subgroups like HER2+
breast cancer with effective targeted therapies are rare and most
cancer drugs are still being applied to 
 large patient populations that include many patients who 
might not respond or benefit.  Also, the response to targeted
agents in human is usually unpredictable.  To address these issues,
we propose SUBA, subgroup-based adaptive designs that simultaneously
search for prognostic subgroups and allocate patients adaptively to
the best subgroup-specific treatments throughout the course of the
trial. The main features of SUBA include the continuous
reclassification of patient subgroups based on a random partition
model and the adaptive allocation of patients to the best treatment
arm based on posterior predictive probabilities.
 We compare the SUBA design with three alternative
designs including 
equal randomization, outcome-adaptive randomization and a design based
on a probit regression.
 In simulation studies we find that SUBA compares favorably
against the alternatives. 

\end{abstract}

\newpage

\section{Introduction}
\label{sec:intro}
\subsection{Targeted Therapy}
With the rapid development in genomics and personalized
medicine it is becoming increasingly more feasible to diagnose
and treat cancer 
based on measurements from genomic interrogations at the molecular
level such as gene expression \citep{van2002gene,snijders2001assembly}, DNA copy numbers \citep{curtis2012genomic, baladandayuthapani2010bayesian}, and epigenetic marks
\citep{wang2008combinatorial, barski2009genomic, mitra2013bayesian}. In particular, pairing genetic traits with targeted
treatment options has been an important focus in recent
research. This has 
led to successful findings such as the use of trastuzumab,
doxorubicin, or taxanes on HER2+ breast cancer
\citep{hudis2007trastuzumab}, and the recommendation against treatment
with EGFR antibodies on KRAS mutated colorectal cancer
\citep{misale2012emergence}.  It is now broadly understood that
patients with the same cancer defined by  classification criteria such
as tumor location, staging, and risk-stratification can respond
differently to the same drug, depending on their genetic profiling.

First proposed by \cite{simon2004evaluating},
``targeted designs'' restrict the eligibility of patients to receive a
treatment based on predicted response using genomic information.
Under fixed sample sizes and comparing to standard equal randomization
with two-arm trials, the authors showed that targeted designs could
drastically increase the study power in situations where the new
treatment benefited only a subset of patients and those patients could
be accurately identified. \cite{sargent2005clinical}
proposed the biomarker-by-treatment interaction design and a 
biomarker-based-strategy design, both using prognostic biomarkers to
facilitate treatment allocations to targeted subgroups.   \cite{maitournam2005efficiency} further showed that the
relative efficiency of target designs depended on (1) the 
relative sizes 
of the treatment effects in biomarker positive and negative subgroups,
(2) the prevalence of the patient group who favorably responds to the
experimental treatment, and (3) the accuracy of the biomarker
evaluation.
Recently, new designs have been proposed by \cite{freidlin2010randomized}, \cite{simon2010clinical} and
 \cite{mandrekar2010predictive}, among others.

BATTLE \citep{kim2011battle} and I-SPY 2 \citep{barker2009spy} are two
widely known biomarker cancer trials using Bayesian designs.  The
design of BATTLE predefined five biomarker groups on the basis of 11
biomarkers, and assigned patients to four drugs using an
outcome-adaptive randomization (AR) scheme.  AR is implemented
with the expectation that an overall higher response rate would be achieved
relative to equal randomization (ER), assuming at least one biomarker
group 
has   variations  in  the   outcome  distributions  across 
 arms. 
 However, the analysis
of the trial data revealed otherwise; the response rate was actually
slightly lower during the AR period than during the initial ER period.
This fact can be attributed to several factors such as possible trends
in the enrolled population, or variations in the procedures for
measuring primary outcomes.  In practice, targeted agents can fail for
reasons such as having no efficacy on the  targeted patients,
being unexpectedly toxic, or uniformly ineffective.  There is a need
for adaptive designs to accommodate the situations above to improve
trial efficiency and maintain trial ethic \citep{yin2012phase,
gu2010simulation, zhu2013adaptive}.
      
 Researchers are  also  developing new designs that allow for the
 redefinition of biomarker groups that could be truly responsive to
 targeted treatments.  
\cite{ruberg2010mean} and 
\cite{foster2011subgroup} developed tree-based algorithms to identify
and evaluate the subgroup effects  by searching the covariate
space for regions with substantially better treatment effects.
Bayesian models are natural candidates for adaptive learning of subgroups, and have
been known and applied in non-medical contexts \citep{loredobayesian2003, kruschkebayesian2008}.  

\subsection{A Subgroup-Based Adaptive Design}
In this paper, we propose a class of SUbgroup-Based Adaptive (SUBA)
designs for targeted therapies which utilize individual biomarker
profiles and clinical outcomes as they become available. 
%

To understand and characterize a clinical trial design it is useful to
distinguish between the patients in the trial versus future patients.
There exist a number of methods that address the optimization for the
patients in the trial.  Most approaches are targeting the optimization
of a pre-selected objective function (criterion). See, for example,
\cite[chapters 8 and 9]{fedorov2013optimal}.
SUBA aims to address both goals, successful treatment of patients in
the trial and optimizing treatment selection for future patients.
We achieve the earlier by allocating each patient on the basis of the
patient's biomarker profile $\bx$ to the treatment with the best currently
estimated success probability. That is, the optimal treatment $t^\ast$
for a patient with biomarker profile $\bx$ is
$$
  t^\ast(\bx) = \arg\max_{t \in \Omega} \hat{\theta}_t(\bx),
$$
where $\hat{\theta}_t(\bx)$ is the posterior predictive response rate 
of a patient with biomarker profile $\bx$ under treatment $t$. 
This can be characterized as a stochastic optimization problem.
In contrast, the optimal treatment selection for future patients is not 
considered as an explicit criterion in SUBA. 
It is indirectly addressed by partitioning the biomarker space into
subsets with different response probabilities for the treatments under
consideration.  Learning about the implied patient subpopulations
facilitates personalized treatment selection for a future patient on
the basis of the patient's biomarker profile $\bx$.
The outcome of SUBA is an estimated partition of the biomarker space
and the corresponding optimal treatment assignments.
\ech

The main assumption underlying the proposed design
approach is that there exist subgroups of patients who
differentially respond to treatments. For example, consider a
scenario with 
two subgroups of patients that respond well to either of two different
treatments, but not both.
 An ideal design should search for such subgroups and link
each subgroup with its corresponding superior treatment.  
That is, a design should aim to
identify subgroups with elevated response rates to particular
treatments. The key innovations of SUBA are that such biomarker subgroups
are continuously redefined based on patients' differential responses
to treatments and that patients are allocated to 
the currently estimated best 
treatment based on posterior predictive inference.  

In summary, 
SUBA conducts subgroup discovery, estimation, and patient allocation
simultaneously. We propose a prior for the partition that classifies
tumor profiles into biomarker subgroups.  The stochastic partition has
the advantage that biomarker subgroups are not fixed up front before
patients accrual.  The goal is to use the data, during the trial, to
learn which partitions are likely to be relevant and could potentially
become clinically useful.    
We define a random partition of tumor profiles using  a
tree-based model that shares similarities with
Bayesian CART algorithms \citep{chipman1998bayesian,
denison1998bayesian}.  We provide closed-form expressions for
posterior computations and describe an algorithm for adaptive patient
allocation during the course of the trial.

\subsection{Motivating Trial}
We consider a breast cancer trial with three candidate
treatments. Patients who are eligible have undergone neoadjuvant
systemic therapy (NST) and surgery. Protein biomarkers for all
patients are measured through biopsy samples by reverse phase protein
arrays (RPPA) at the end of NST, but before surgery. The first
treatment is a poly (ADP-ribose) polymerase (PARP) inhibitor, which
affects  
DNA repair and cell death programming.  
The second treatment is a PI3K pathway inhibitor, which
affects cell growth, proliferation, cell differentiation and
ultimate survival. The
third treatment is a cell cycle inhibitor that targets the cell cycle
pathway. The main goal is to identify
for each of the three treatments subgroups of patients that will
respond favorably to the respective treatment. 

The paper proceeds as follows. Section \ref{sec:model} presents the
probability model of SUBA design and computation details for
implementing the design. Section \ref{sec:simu} examines the operating
characteristics based on simulation studies. We conclude with a brief
discussion in Section \ref{sec:discussion}.

\section{Methodology}
\label{sec:model}
\subsection{Sampling Model}
\label{sec:py}
Assume that $T$ candidate treatments 
are  under consideration in a clinical trial.
We use $t \in \Omega = \{1, \ldots, T\}$ to index the treatments and
$i=1,\ldots,N$ to index patients. We assume a maximum sample size of
$N$ patients.
The primary
outcome for each patient is a binary variable $y_i \in \{0,1\}$.
We assume that $y_i$ can be measured without
delay. 
We denote with $\bx_i=(x_{i2}, \dots, x_{iK})'$ the biomarker profile
of the $i$-th patient, recorded at baseline. 
We assume that all biomarkers $x_{ik}$ are continous, $x_{ik} \in
\Re$. 
Finally, let $z_i$  denote the treatment  allocation for patient $i$
 with $z_i=t$ if patient $i$ is assigned to treatment $t$.

The underlying assumption of a biomarker clinical trial is that there
exist subgroups of patients that differentially respond to the same
treatment. 
For example, subgroup $1$ may respond well to treatment $t_1$ but
not $t_2$ while subgroup $2$ may respond well to treatment
$t_2$ but not $t_1$. However, the
subgroups are not known before the trial and must be estimated
adaptively based on response data and biomarker measurements from
 already  treated patients.  To estimate the subgroups and their
expected response rates to treatments, we propose a random partition
model.  Assuming that all $K$ biomarker measurements are continuous,
$x_{ik} \in \Re$, 
we construct patient subgroups by defining 
a partition of the biomarker space $\Re^K$.
A partition is a family of subsets 
$\bPi=\{S_1, S_2, \dots, S_M\}$, where $M$ 
is the size of the partition and $S_m$ are the 
partitioning subsets such that
$S_m \cap S_l = \emptyset$ and $\cup_{m} S_m = \Re^K$. 
The partition of the biomarker sample space implies a partition
of the patients into biomarker subgroups. 
Patient $i$ belongs to biomarker subgroup $m$ if  $\bx_i\in S_m$. 
We will construct a prior probability measure for $\bPi$ in the next
section. In the following discussion we will occasionally refer
to $S_m$ as a subset of patients, implying the subset of patients that
is defined by the partitioning subset $S_m$. 

We define a sampling model for $y_i$ conditional on
$\bx_i$ and $\bPi$ as
\begin{equation}
  p(y_i=1 \mid z_i=t, \bPi, \bx_i \in S_m) = \th_{t,m},
\label{eq:pyi}
\end{equation}
where $\theta_{t,m}$ is the response rate of treatment $t$ for
a patient in subgroup $S_m$. Thus the joint likelihood function for $n$
patients is the product of $n$ such Bernoulli probabilities, 
using $\th_{t,m}$ and $(1-\th_{t,m})$ depending on the recorded
outcomes $y_i$.
In each biomarker subgroup $S_m$, let
$n_m = \sum_{i} I(\bx_i \in S_m)$ count the number of patients,
$n_{mt} = \sum_i I(\bx_i \in S_m,\; z_i = t)$ the number of patients
assigned to treatment $t$, and
$n_{mty}=\sum_i I(\bx_i \in S_m,\; z_i = t,\; y_i = y)$  the number of
patients in group $m$ assigned to $t$ with response $y_i=y$.
Here $I(\cdot)$ is the indicator function. 
Let $\byn=(y_1, \dots, y_n)'$,
$\bXn=\{\bx_i\}_{i=1}^n$, $\bzn=(z_1, \dots, z_n)'$, and
$\btheta=\{\theta_{t,m};\; t=1,\ldots, T,\;m=1,\ldots,M\}$.  
Then 
$$
  p(\byn \mid \bXn, \bzn, \bth, \Pir) =
  \prod_m\prod_t
    \theta_{t,m}^{n_{mt1}}(1-\theta_{t, m})^{n_{mt0}}.
$$
Adding a prior on $\Pi$ and $\bth$ we complete \eqref{eq:pyi} to 
define a 3-level hierarchical model 
\begin{eqnarray}
  \label{eq:model}
   p(\byn, \btheta, \bPi\mid \bXn, \bzn)\propto 
   p(\byn\mid \bXn, \bzn, \btheta, \bPi)\, p(\btheta\mid \bPi)\, p(\bPi).
\end{eqnarray}
The last two factors define the prior model for $\btheta$ and $\bPi$. 
We assume $\theta_{t,m}\mid \bPi
\stackrel{i.i.d}{\sim} \mathrm{Beta}(a, b)$ and discuss the prior for
$\bPi$ next. Posterior inference on $\Pi$ and $\bth$ provides
learning on subgroups and their treatment-specific response rates.
Posterior probabilities for $\Pi$ and $\bth$ are the key inference
summaries that we will later use to define the desired adaptive trial
design. 

\subsection{Random Biomarker Partition $\Pi$}
We propose a tree-type random partition $\bPi$ on the biomarker space
$\Re^K$ to define random biomarker subgroups.  A partition is obtained
through a tree of recursive binary splits.  Each node of the tree
corresponds to a subset of $\Re^K$,
and is either a final leaf which defines one of the partioning
subsets $S_m$, or it is in turn split into two descendants. 
In the latter case the two descendants are defined by
first selecting a biomarker $k$ and then splitting the current
subset by thresholding $x_{ik}$.
The threshold  splits the  ancestor  set      into   two  components.  
A sequence of such splits
generates a partition of $\Re^K$ as the collection of the
resulting subsets.  
For the motivating breast cancer trial, we limit the partition to
 at most eight biomarker subgroups in the random partition. 
 We impose this constraint to limit the number of subgroups with 
critically small numbers of patients, 
and therefore only allow three rounds of random splits.

\begin{figure}[h]
\centering

\includegraphics[scale=0.8]{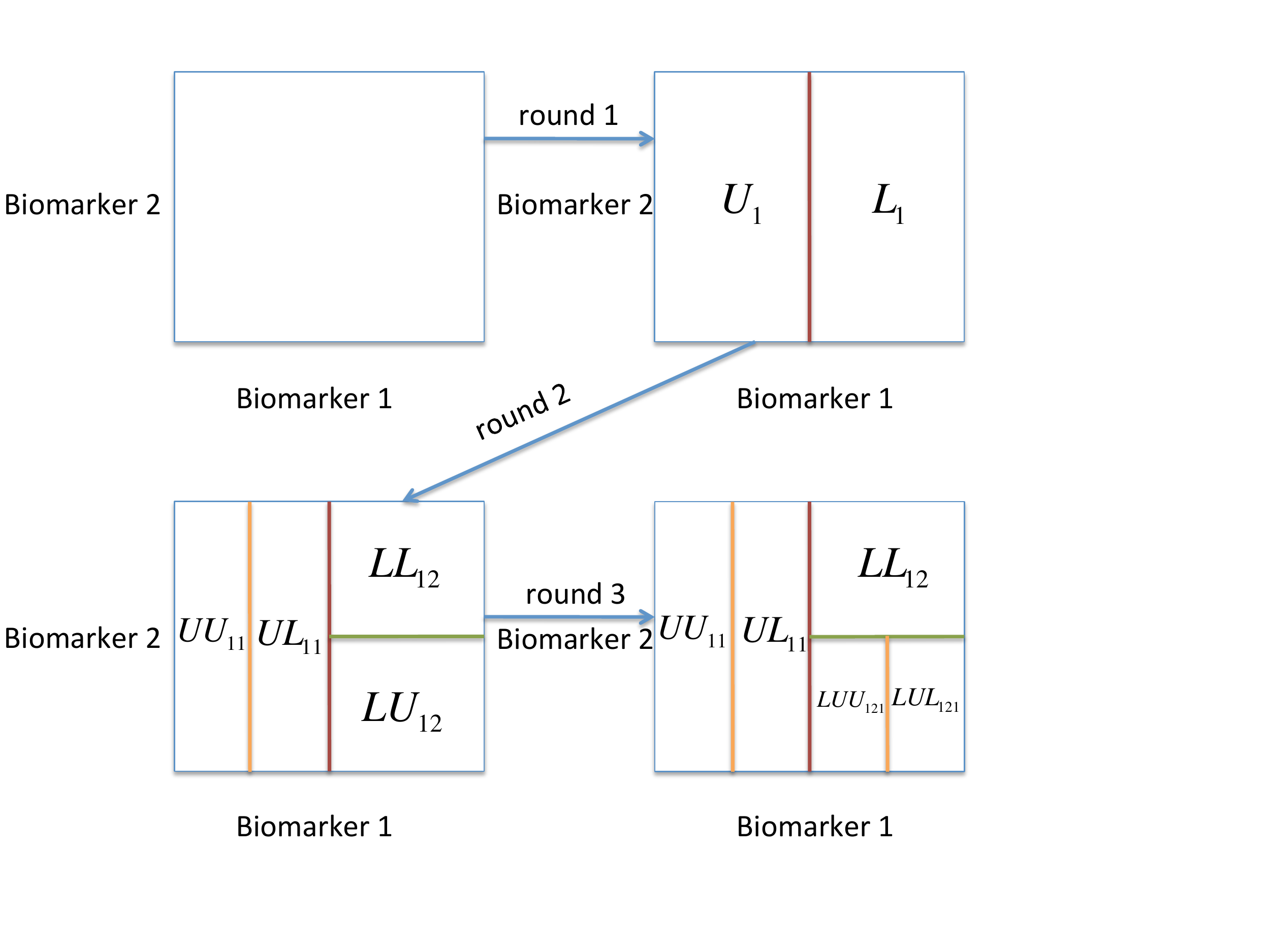}
\caption{An illustration of $p(\Pi)$ with three rounds splits. The example shows that with three rounds of split, the initial space of two biomarkers is partitioned into five sets $\{UU_{11}, UL_{11}, LL_{12}, LUU_{121}, LUL_{121}\}$.}
\label{fig:partition}
\end{figure}

An example is shown in
Figure \ref{fig:partition}.  The figure shows a realization of the
random partition with $K=2$ biomarkers.
In each round, we consider each of the current subsets and
either do not split it further with probability
$v_0$ or with probability $v_k$ choose biomarker $k$ to split the
subset into two parts. 
If an  ancestor  subset $S$ is split  by  the  $k$-th  biomarker,
then the resulting partition contains two new subsets, 
defined by $\{i: x_{ik} \ge med_k(S)\} $ and $\{i: x_{ik} < med_k(S)\}$,  where $med_k(S)$  is the median of  
$x_{ik}$ and is computed across all  available  data points in the subset  $S$.
  That is, $med_k(S)$ is a conditional median which
can  vary   during  the  course   of  the  trial,  as  more   data  become  available.
     %
In Figure \ref{fig:partition} the sequence of splits
is as follows. We first split on $x_{i1}$. In the second round the two
resulting subsets are split on $x_{i1}$ and $x_{i2}$, respectively.
 In a third round of splits, only one subset of the earlier four subsets is split on $x_{i1}$ again, three others are not
further split. 

 Let $\cPi$ be the sample space of all possible partitions based on
the three rounds of splits. For each partition $\Pi \in \cPi$, we
calculate the prior probability $\pPir$ based on the above random
splitting rules. For example, the partition $\Pi$ in Figure
\ref{fig:partition} has prior probability
 \begin{equation} 
\label{eq:priorPio}
   p(\Pi) \propto 
   v_1 \times
   v_1 \, v_2 \times 
   v_0 \, v_0\, v_0 \, v_1,  
\end{equation}   
with the three factors corresponding to the three rounds of splits.

We  use a \ech variation of  the  described  probability model.    
The  main  rational  is  that,   if a biomarker   is selected  for  an  initial  split, then  it 
is  desirable  to  augment  the   probability of  splitting it  again  at  the  subsequent  levels  in the tree.
The  goal  is  to  facilitate  the identification of   relevant subgroups  maintaining   the  simplicity  of  the  partition  model.  
To 
implement this, in each possible partition $\Pi$, we calculate
$K$ as the number of distinct biomarkers  selected  in the three rounds
of splits. We then add an additional penalty term proportional  to $\phi^{K}$ to
the above  prior probability of $\Pi$, so that the prior favors partitions that repeatedly split on the same
marker. 
For example, in Figure \ref{fig:partition}, the 
modified prior probability is
\begin{equation} 
  p(\Pi) \propto v_1^3\, v_2\, v_0^3 \times \phi^{2}. \label{eq:priorPi}
\end{equation} 
Similarly, we can calculate the prior probability for
any partition $\Pi$  in $\cPi$.
When  $\phi=1$  the  two  probability    models  that  we  described   coincide  while   values  of $\phi$   
in $(0,1)$ allow one   to   tune  the   concentration  of   over  partitions   that   split  over  a parsimonious   number  of  biomarkers.

\subsection{ Decision Rule for Patient Allocation} 
\label{sec:SUBA} 
A major objective of the SUBA design is to assign
future patients to superior treatments based on their biomarker
profiles and the observed outcomes of all previous patients. Assuming
that the outcomes of the first $n$ patients have been observed, we
denote by $q(t, \xno)$ the posterior predictive probability of
response under treatment $t$ for an  $(n+1)^{th}$ patient
with biomarker profile $\xno$.  Denoting the observed trial data
$\Dn = \{\byn, \bXn, \bzn\}$, based on \eqref{eq:model},
\begin{multline}
  q(t, \xno) \equiv \Pr(y_{n+1}=1\mid \xno,  z_{n+1}=t, \Dn)\\
  = \sum_{\Pi_r \in \cPi} 
   \Pr(y_{n+1}=1 \mid \xno, z_{n+1}=t, \Pir, \Dn)\;  
             \Pr(\Pir \mid \Dn). 
  \label{eq:pred}
\end{multline}
The posterior probability $\Pr(\Pi_r \mid \Dn)$  
can be computed as follows.  Given a partition $\Pi_r=(S_1, \dots, S_{M_r})
\in \cPi$, all $n$ patients are divided into $M_r$ biomarker
subgroups.  
Recall the definition of $n_m, n_{mt}$ and $n_{mty}$ from Section 
\ref{sec:py}. 
The posterior distribution of $\Pi_r$ is 
$$
  \Pr(\Pir\mid \Dn) \propto   \Pr(\Pir)\, p(\Dn \mid \Pir) =
  \Pr(\Pir) \,
  \prod_m
  \prod_t
  \left\{\int \prod_{\bx_i \in S_m}p(y_i\mid \bx_i, z_i=t, \th_{t,m})
         \; dp(\th_{t,m}) \right\},
$$
where $\Pr(\Pir)$ is the prior probability of partition $\Pi_r$ that
can be calculated as in \eqref{eq:priorPi}. 
Let $B(a,b)=\Gamma(a)\Gamma(b)/\Gamma(a+b)$ denote the beta
function, and let $\Be(x;\, a,b) \propto x^{a-1}(1-x)^{b-1}$ denote a beta p.d.f.
With  independent  $\Be(x;\, a,b)$  prior  distributions  for  the  $\th_{t,m}$  parameters we can further simplify the  above  equation  to 
\begin{multline}
  \Pr(\Pir\mid \Dn) \propto
  \Pr(\Pir)\,
  \prod_m\prod_t\left\{\int
    \theta_{t,m}^{n_{mt1}}(1-\theta_{t, m})^{n_{mt0}}\;
    \Be(\th_{t,m};\, a,b) ~d\theta_{t,m}\right\} \, =\\
  = \Pr(\Pir)\; 
  \prod_m\prod_t
    \frac{B(a+n_{mt1},b+n_{mt0})}{B(a,b)}.
  \label{eq:pos1}
\end{multline}
The conditional probability 
$\Pr(y_{n+1}=1 \mid \xno, z_{n+1}=t, \bPi, \Dn)$
is the integral of \eqref{eq:pyi} with respect to the
$\Be(a+n_{mt1},b+n_{mt0})$ posterior on $\th_{t,m}$.
Then 
\begin{eqnarray}
\Pr(y_{n+1}=1 \mid \xno, z_{n+1}=t, \Pi_r, \Dn)  &=&
 \sum_{m} I(\xno \in S_m)\int \th_{t,m}\,
  dp(\theta_{t,m} \mid \Pi_r, \Dn)  \nonumber\\
  &=&\sum_m I(\xno \in S_m) \frac{a+n_{mt1}}{a+b+n_{mt}}.
\label{eq:pos2}
\end{eqnarray}
Let $m(\xno,\Pi)$ index the partitioning subset with $\xno \in S_{m(\xno,\Pi)}$. The sum over $m$ in (\ref{eq:pos2}) reduces to just the term with $m=m(\xno,\Pi)$. 
Combining \ech (\ref{eq:pos1}) and (\ref{eq:pos2}), we compute the
posterior predictive response rate of $(n+1)^{th}$ patient receiving
treatment $t$ in closed form 
\begin{equation}
  q(t, \xno) =
  \sum_{\Pi_r \in \cPi}\Pr(y_{n+1}=1\mid \xno, z_{n+1}=t,\Pir,\Dn)\;
  p(\Pir\mid \Dn).
\label{eq:predictive}
\end{equation}

Denote with $\zs_{n+1} \in \Omega$ the treatment decision for the $(n+1)^{th}$
patient. We choose  $\zs_{n+1}$ by adopting a minimum posterior
predictive loss approach  described in 
\cite{gelfand1998model}. 
Under a variety of loss functions (such as the 0-1 loss), the optimal rule that minimizes the posterior
predictive loss is 
\begin{equation}
\zs_{n+1}=\arg\max_{t\in \Omega} \; q(t,\xno).
\label{eq:allo}
\end{equation}
See  \cite{raiffa1961applied} or 
\cite{gelfand1998model} for details.
Alternatively, one could use the probabilities $q(t, \xno)$ in a
biased randomization $p(\zs_{n+1}=t)\propto q(t, \xno)^c$,
as proposed in  
\cite{thall2007practical}.

\subsection{The SUBA Design}
Computing the posterior predictive response rates for all candidate
treatments allows us to compare  treatments and monitor the trial
accordingly. If one treatment is inferior to all other
treatments, that treatment should be dropped from the trial.  If there
is only one treatment left after dropping inferior treatments, the
trial should be stopped early  due to ethical and logistics reasons.

The SUBA design starts a trial with a run-in phase during 
which patients are equally randomized to treatments. After
the initial run-in, we continuously monitor the trial 
until either the trial is stopped early based on a stopping rule, or the trial is stopped after reaching a prespecified maximum
sample size $N$.

We include rules to exclude inferior treatments and
stop the trial early  if indicated. Recall
that the biomarker space is $\Re^K$. 
Consider the $k$-th biomarker and observed biomarker values $\{x_{1k},
\ldots, x_{nk}\}$. 
We define an equally spaced grid of size $H_0$
between $\min_k$ and $\max_k$, 
where $\min_k$ and $\max_k$ are the observed smallest and largest
values for that biomarker. 
Taking the Cartesian product of these grids we 
then create a $K-$dimensional grid $\tbx$ of size $H = H_0^K$.
Let $\tx_h \in \Re^K$, $h=1,\ldots,H$, denote the list of all grid
points. 
After an initial run-in phase with equal randomization, we evaluate
the posterior predictive response rate $q(t, \tx_h)$ for treatment $t$
for each $\tx_h$. 
Any treatment $t^\star$ with uniformly inferior success probability $$
   q(t^*, \tx_h) <    q(t,\tx_h ),
   \mbox{ for all } h=1, \ldots, H \mbox{ and } t\neq t^\star
$$  
 is \ech dropped from the trial.
That is, we remove $t^\star$ from the list of treatments,
$\Omega \equiv \Omega\setminus \{t^\star\}$.
Also, if  only one treatment is left in the trial,  then
the trial is stopped early.

Alternatively to the construction of the grid $\tbx$, any available
data set of typical biomarker values $\tx_h \in \Re^K$ could be used.
For large $K$ this is clearly preferable.
If such data were available, it could also be used for an alternative
definition of $med_k$ in the specification of the splits in the prior
for $\Pir$ discussed earlier.

\bigskip

The SUBA design consists of the following steps.
\begin{center}
\fbox{\fbox{\parbox{16cm}{

\begin{description}
\item[1. Initial run-in.] Start the trial and randomize $n < N$ patients
    equally  to $T$ treatments in the set $\Omega$.
\item[2. Treatment exclusion and early stopping. \ech] Drop treatment $t^*$
  if $q(t^*, \tx_h)<q(t,\tx_h)$ for all $t\neq t^*$ and $h=1,
  \ldots, H$. Set $\Omega = \Omega \setminus \{t^*\}$.
  If  enrollment   remains  active  only for  a  single  treatment  $t$  then
  stop the trial.  
\item[3. Adaptive patient allocation.] 
  Allocate patient $(n+1)$ to treatment $\zs_{n+1}$ according to
  \eqref{eq:allo}. 
  When the response $y_{n+1}$ is available, go back
  to step 2 and repeat for patients 
  $n+2, n+3, \dots, N$.
\item[4. Reporting patient subpopulations.]
  Upon conclusion of the trial we report the estimated partition
  $\Pi$ together with the estimated optimal treatment allocations.
\end{description}
}}}
\end{center}

In step 4, summarizing  the posterior  distribution over random
partitions and determining the best partition over a large number of possible
partitions $\Pi $   is a challenging problem. 
Following 
\cite{medvedovic2004bayesian} we define an $(N \times N)$
association matrix $G^{\Pi_r}$ of co-clustering indicators 
for each partition $\Pi_r$.
Here $G^{\Pi_r}_{ij}$ is 
an indicator of patients $i$ and $j$ being in the same subgroup
with respect to the biomarker partition $\Pi_r$.
\cite{dahl2006model} introduced a least-squares estimate for random
partitions using draws from Markov chain Monte Carlo (MCMC)
posterior simulation.
Following their idea,  we propose a least-square summary
$$
   \Pi^{LS}=\arg min_{\Pi_r} ||G^{\Pi_r}-\hat{G}||^2,
 $$
where $\hat{G}=\sum_{r}G^{\Pi_r}\Pr(\Pir\mid \Dn)$ 
is the posterior mean association matrix
and  $||A||^2$ denotes  the sum of squared elements  of a  matrix $A$.
In words,  $\Pi^{LS}$ minimizes the sum of squared
deviations of between an association matrix $G^{\Pi_r}$ and the
posterior mean $\hat{G}$.  

Alternatively one could report a partition that minimizes the average
squared deviation, averaging with respect to $p(\Pi_r \mid \mathcal{D})$.  That is, 
minimize posterior mean squared distance instead of squared distance to the
posterior mean association matrix.
While the earlier has an appealing justification as a formal Bays
rule, the latter is easier to compute.

\section{Simulation Studies}
\label{sec:simu}
\subsection{Simulation Setup}

We conduct simulation studies
to evaluate the proposed design. 
 The setup is chosen to mimic the motivating breast cancer study.
For each simulated trial, we  fix a maximum sample size of 
$N=300$ patients in a three-arm study with three treatments $t=1,2,3$.
We assume that a set of $K=4$ biomarkers are measured
at baseline for each patient and generate $x_{ik}$
from a uniform distribution on $[-1,1]$, i.e.,
$x_{ik} \sim \Unif \ (-1, 1)$. 
The hyperprior parameters are fixed as 
$v_k=1/(K+1)$, $k=0, 1, \dots,
K$, $\phi=0.5$, $a=1$ and $b=1$. That is, each biomarker has the
same prior probability of being selected for a split, 
and the response rates $\th_{t,m}$ have uniform priors.
To set up the grid $\tbx$ for the stopping rule
we select $H_0=10$ equally spaced points on each biomarker
subspace, and thus $H=10,000$ grid points in $\tbx$.
During the initial run-in phase, $n=100$ patients are equally randomized to three treatments. 

\begin{figure}[h]
\includegraphics[scale=0.7]{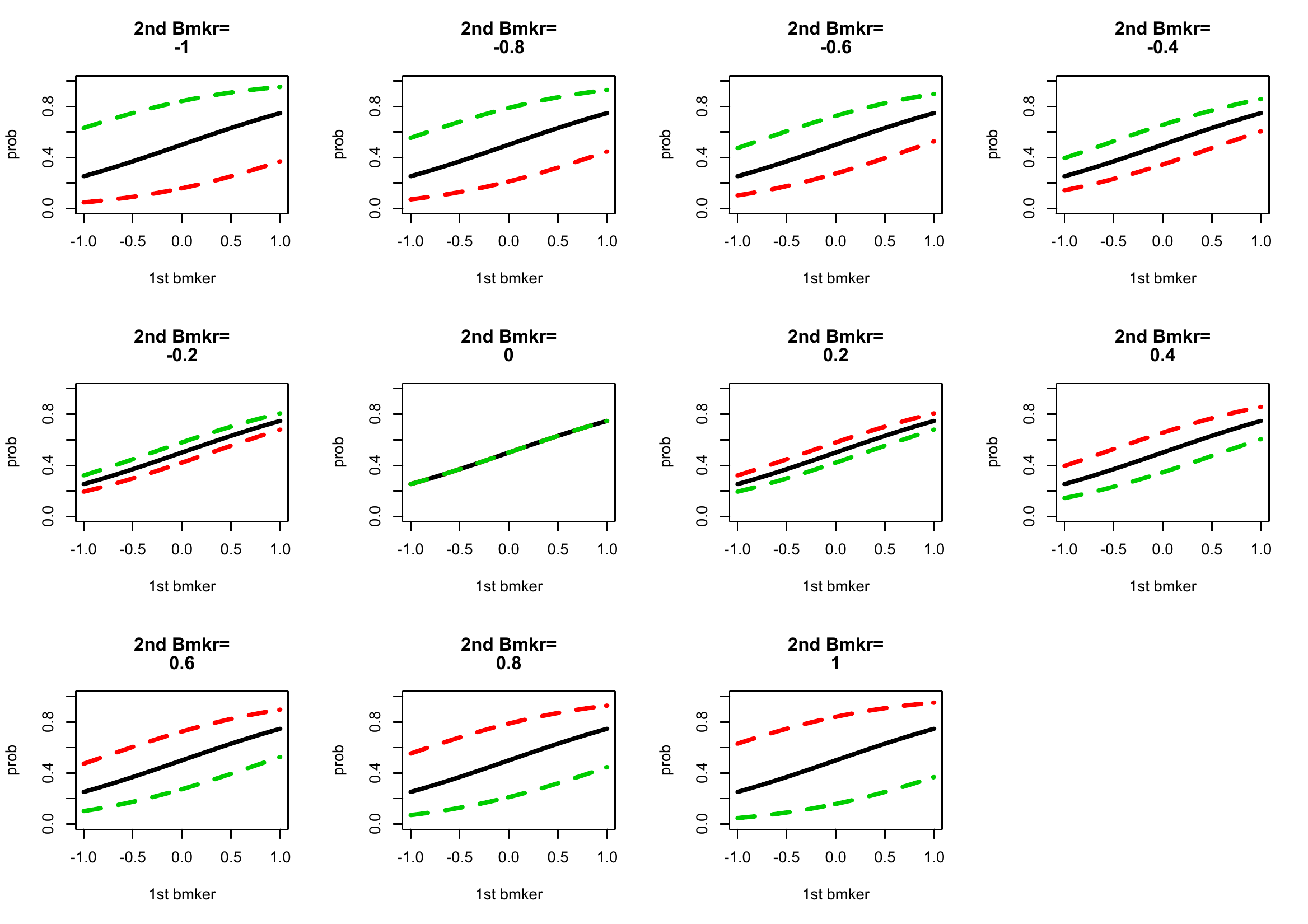}
\caption{Display of Scenario 2. \jj The probabilities of response versus the measurements of the first biomarker given fixed values of the second biomarker. Red, black and green lines represent three treatments 1, 2 and 3 respectively. }
\label{fig:simutruth}
\end{figure}

\paragraph*{Scenarios 1 through 6.}
We consider six scenarios and simulated $1,000$ trials for each
scenario. In the first two scenarios, we assume that biomarkers
$x_{i1}$ and $x_{i2}$ are relevant to the response, but not biomarkers
$x_{i3}$ and $x_{i4}$. 
The simulation truth for the outcome $y_i$ is a probit
regression.  
Specifically, we assume that the true response rates  for a patient with
covariate vector $\bx_i$ under  treatments 1, 2 or 3
are $\theta_{1i}=\Phi_{\mu=0, \sigma=1.5}(x_{i1}+1.5x_{i2})$,
$\theta_{2i}=\Phi_{\mu=0, \sigma=1.5}(x_{i1})$, or
$\theta_{3i}=\Phi_{\mu=0, \sigma=1.5}(x_{i1}-1.5x_{i2})$,
respectively, where $\Phi_{\mu=0, \sigma=1.5}$ is the cumulative
distribution function (CDF) of a Gaussian distribution with $\mu=0$
and $\sigma=1.5$. 
Figure \ref{fig:simutruth} plots the response rates
under three treatments versus  $x_{i1}$ 
given different values of $x_{i2}$. 
The red lines represent treatment 1, black lines  refer  to treatment 2 and
green lines   to treatment 3. 
Treatment 3 is always the most effective arm when  $x_{i2}<0$, 
the three treatments have equal success rates when $x_{i2}=0$, 
and treatment 1 is superior when $x_{i2}>0$.
In summary, the optimal treatment is a function of 
the second biomarker, $x_{i2}$. That is, $x_{i2}$  identifies  the
optimal treatment selection. 
The response rates of three treatments increase with $x_{i1}$,
but the ordering of the three treatments does not change   varying  the  first  biomarker.
Therefore, $x_{i1}$ is only predictive of response, but   ideally  should  not  be  involved    for  treatment  selection. 
To assess the performance of SUBA under this setup, we select two
scenarios. In an over-simplified scenario 1, we assume that all the
patients have $x_{i2} = 0.8$.
Thus, treatment 1 is more effective than 2, which in turn is more effective
than 3.  In scenario 2, we do not fix the values of $x_{i2}$ and
randomly generate all biomarker values.

\begin{figure}[h]
\centering
\begin{tabular}{cc}
\includegraphics[scale=0.45]{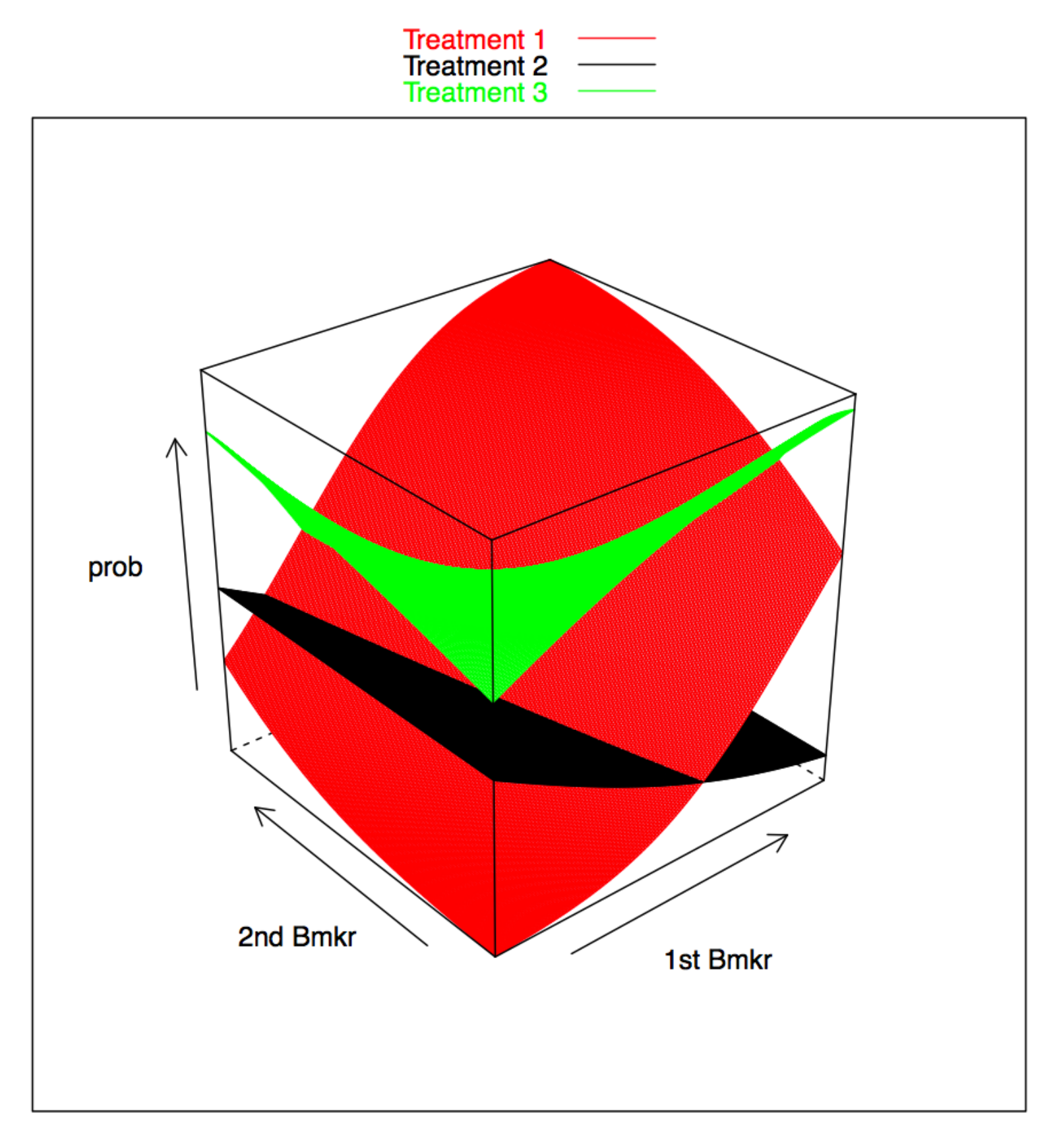}&\includegraphics[scale=0.45]{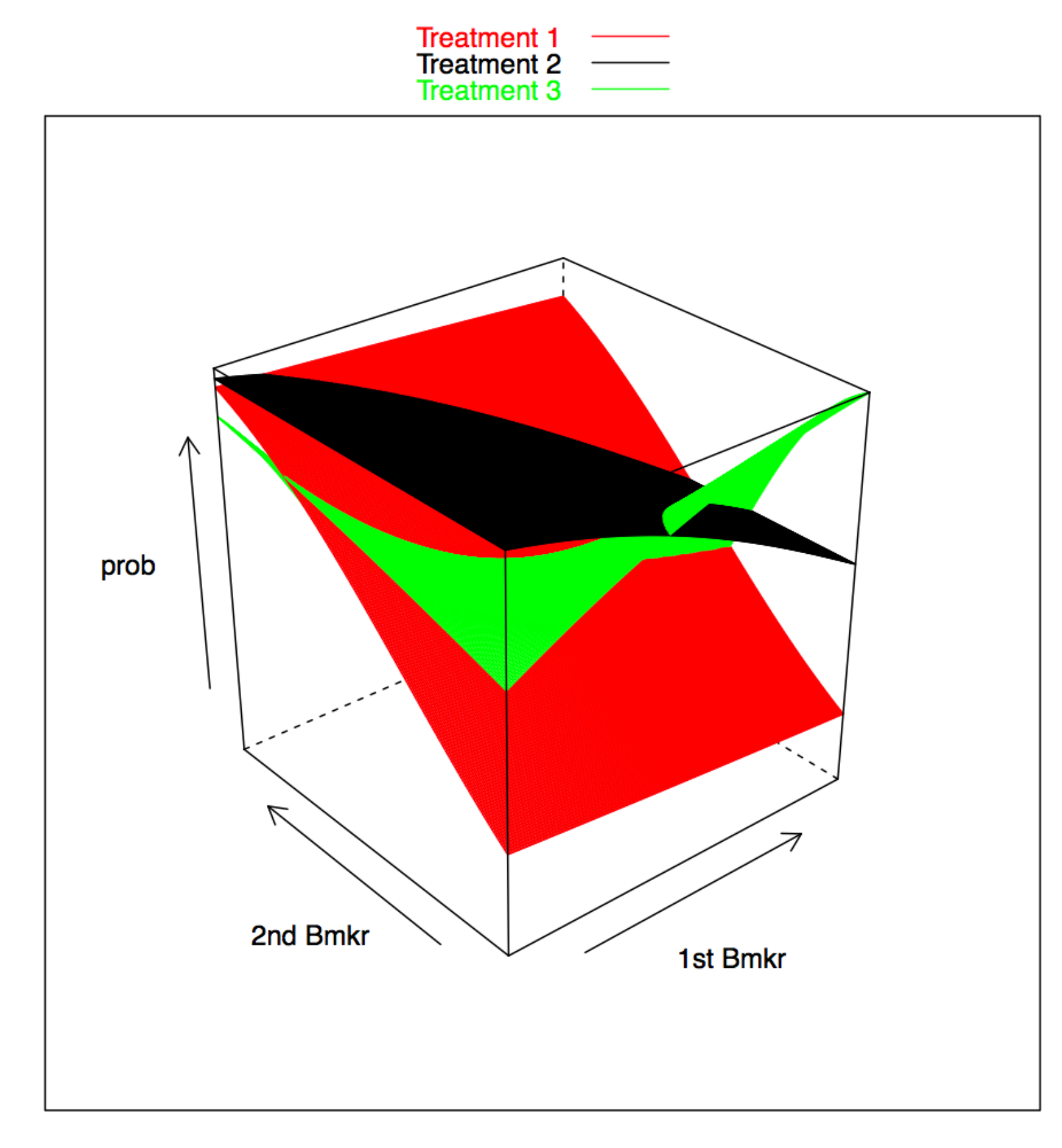}\\
(a) When 3rd biomarker=0.6&(b) When 3rd biomarker=-0.6 \\
\end{tabular}
\caption{Display of Scenario 3. The probabilities of response versus the measurements of the first and the second biomarkers given the fixed values of the third biomarker at 0.6 (a) and -0.6 (b). Red, black and green lines represent three treatments 1, 2 and 3 respectively. }
\label{fig:simutruth3}
\end{figure}

In scenario 3, we assume that biomarkers 1, 2 and 3 are related to
the response and there are interactions. The true response rates 
under
treatments 1, 2, or 3 are
$\theta_{1i}=\Phi_{\mu=0,
\sigma=1.5}(x_{i1}+1.5x_{i2}-0.5x_{i3}+2x_{i1}x_{i3})$,
$\theta_{2i}=\Phi_{\mu=0, \sigma=1.5}(-x_{i1}-2x_{i3})$, or
$\theta_{3i}=\Phi_{\mu=0,
\sigma=1.5}(x_{i1}-1.5x_{i2}-2x_{i1}x_{i2})$, respectively. Figure
\ref{fig:simutruth3} plots the   
response rates under three treatments versus $(x_{i1},x_{i2})$
given $x_{i3}=0.6$ (Figure
\ref{fig:simutruth3}a) and given $x_{i3}= -0.6$ (Figure
\ref{fig:simutruth3}b). 
Here, all three markers are predictive of the ordering of the
treatment effects in a complicated fashion.

We design scenarios 4 and 5 with 
treatment 3 being uniformly inferior to 
treatments 1 and 2.
We assume that the response rates under treatments 1 and 2
are $\theta_{1i}=\Phi_{\mu=0,
  \sigma=1.5}(x_{i1}^2/2+x_{i1}x_{i2}/2)$ or
$\theta_{2i}=\Phi_{\mu=0, \sigma=1.5}(x_{i2}^2/2-x_{i1}x_{i2}/2)$. 
The implied minimum response rate for treatments
1 and 2 is 0.37 and the response rates of treatment 1 and 2 are 
close for all biomarker values (differences range from -0.24 to 0.24
with the first quantile  across  biomarker profiles equal to
-0.06 and the third quantile   equal to  0.09).  We
assume $\theta_{3i}=0.15$ in scenario 4 and $\theta_{3i}=0.3$ in
scenario 5, thus $\theta_{3i}\leq \min(\theta_{1i}, \theta_{2i})$ for
all $x_{i1}$ and $x_{i2}$. 
So we can expect that treatment 3 should be excluded in both scenarios.

Finally, Scenario 6 is a null case, in which no biomarkers are
related to response. We assume that the response rates under the three
treatments for all the patients are the same at 40\%,
that is, $\theta_{1i}=\theta_{2i}=\theta_{3i}=0.4$.  

\paragraph*{Comparison.}
For comparison, we implement a standard design with equal
randomization (ER), an outcome-adaptive randomization (AR) design, and
a 
design based on a probit regression model (Reg).
 In the ER design, all patients are equally randomized to the three treatments and their responses are generated from $\mathrm{Bernoulli}(\theta_{ti})$ for patient $i$ receiving treatment $t$, $t=1, 2, 3$ and $i=1, \dots, N$. The values of $\theta_{ti}$ are defined  by  the Gaussian CDFs given above.
Under the AR design, we assume that three predefined biomarker
 subgroups are fixed before the trial (similar to the BATTLE    trial
 \cite{kim2011battle}). 
We assume that the three
 subgroups are defined as $\{x_{i1}<-0.5\}$, $ \{-0.5\leq x_{i1}\leq
 0.5\}$ and $\{x_{i1}>0.5\}$,
using the quartiles of the empirical distribution of biomarker
$x_{i1}$ as thresholds. 
Apparently, these subgroups are wrongly
 defined and do not match the true response curves in scenarios
 1-6. The mismatch is deliberately chosen to evaluate the
 importance of correctly defining subgroups. 
 Let $p_{tb}$ be the
 response rate of treatment $t$ in subgroup $b$, and $n_{tb}$ the
 total number of patients receiving treatment $t$ in subgroup $b$,
 $t=1, 2, 3$ and $b=1, 2, 3$. For  this  design  we  use  the  model $y_i \mid x_i\in b\sim
 \mathrm{Binomial}(n_{tb}, p_{tb})$. With a
 conjugate beta prior distribution beta(1,1) on $p_{tb}$, we easily
 compute the posterior of $p_{tb}$ as $p_{tb}\mid \mathcal{D}\sim
 \mathrm{beta}(n_{tb1}+1, n_{tb}-n_{tb1}+1)$,  where $n_{tb1}$ is the
 number of patients who responded to treatment $t$ in subgroup
 $b$. Then under the AR design, we first equally randomize 100
 patents to the three treatments, and adaptively randomize the next
 200 patents sequentially.   The AR probability for a future patient
 in subgroup $b$ equal
 $\hat{p}_{tb}/(\hat{p}_{1b}+\hat{p}_{2b}+\hat{p}_{3b})$, where
 $\hat{p}_{tb}$ is the posterior mean $(n_{tb1}+1)/(n_{tb}+2)$,   alternatively     other  summaries  of  the $(p_{1b},p_{2b},p_{3b})$  posterior  can  be  used to  adapt  treatment  assignment  \cite{thall2007practical}. 
Under the Reg design, we model binary outcomes using a probit
 regression. In the probit model, the inverse standard normal CDF of
 the response rate is modeled as a linear combination of the
 biomarkers and treatment,  $\Pr(y_i=1\mid z_i, \bx_i) =
 \Phi(\beta_0z_i+{\bm \beta}'_1\bx_i)$. The parameters $\beta_0$ and
 ${\bm \beta}_1=(\beta_{11}, \ldots, \beta_{1k})$ are obtained using maximum likelihood estimation. 
Under the Reg design, we randomize the first 100 patients with
equal probabilities to the 
three treatments, and then assign the next 200 patients
to the treatment with estimated best success
probability, sequentially.

\subsection{Simulation Results}
\begin{figure}[h]
\centering
\begin{tabular}{ccc}
\includegraphics[scale=0.3]{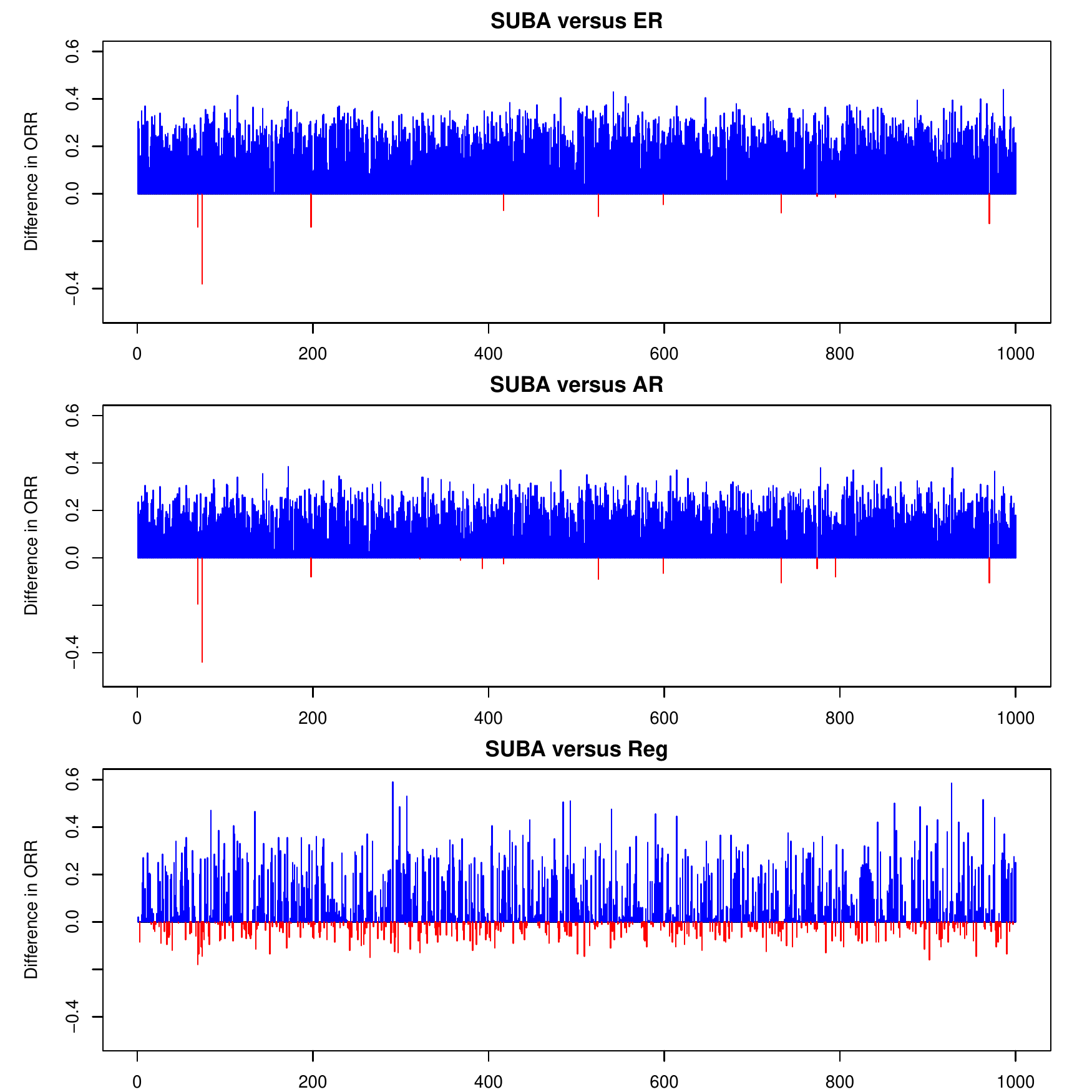}&\includegraphics[scale=0.3]{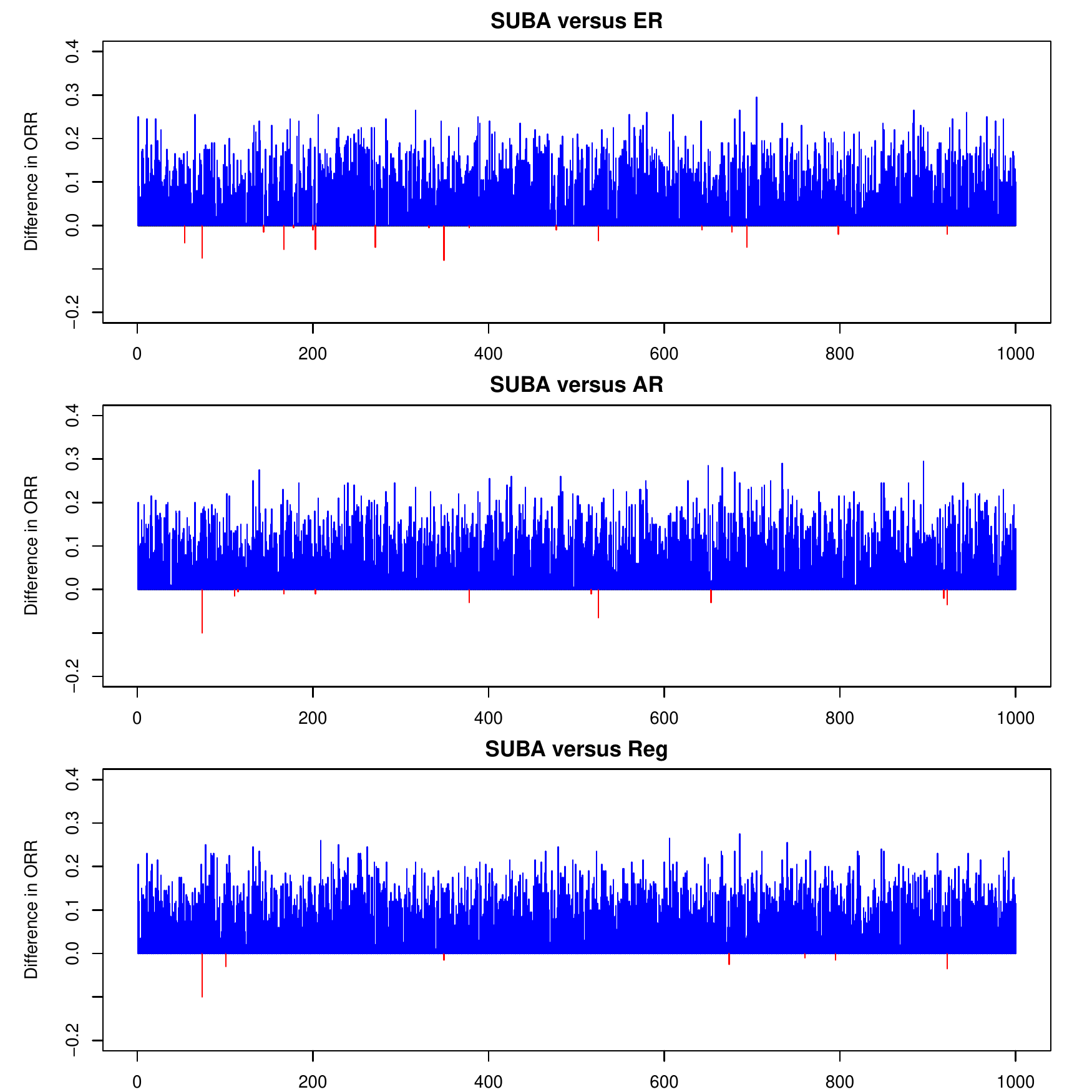}&\includegraphics[scale=0.3]{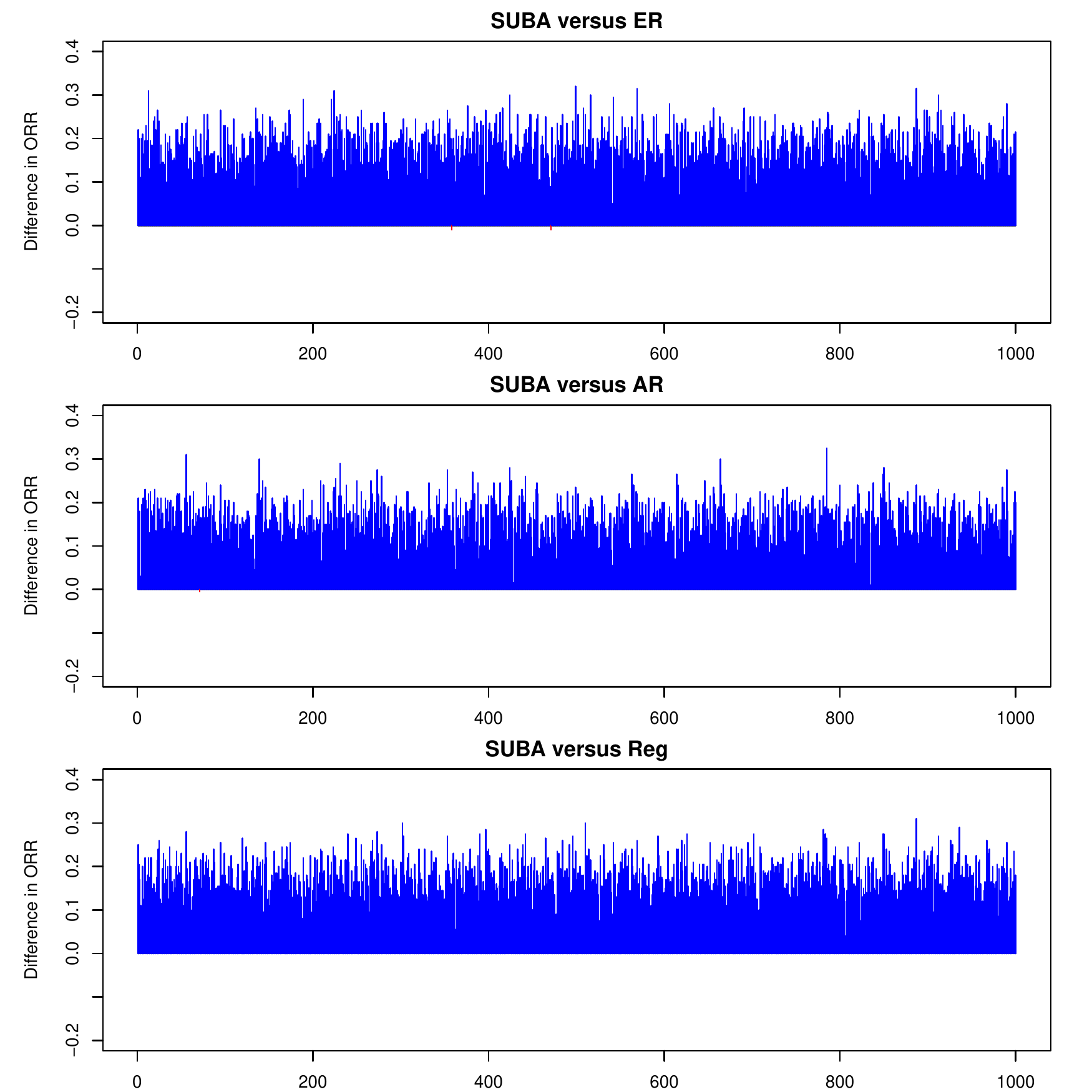}\\
\vspace{.4in}
Scenario 1& Scenario 2& Scenario 3\\
\includegraphics[scale=0.3]{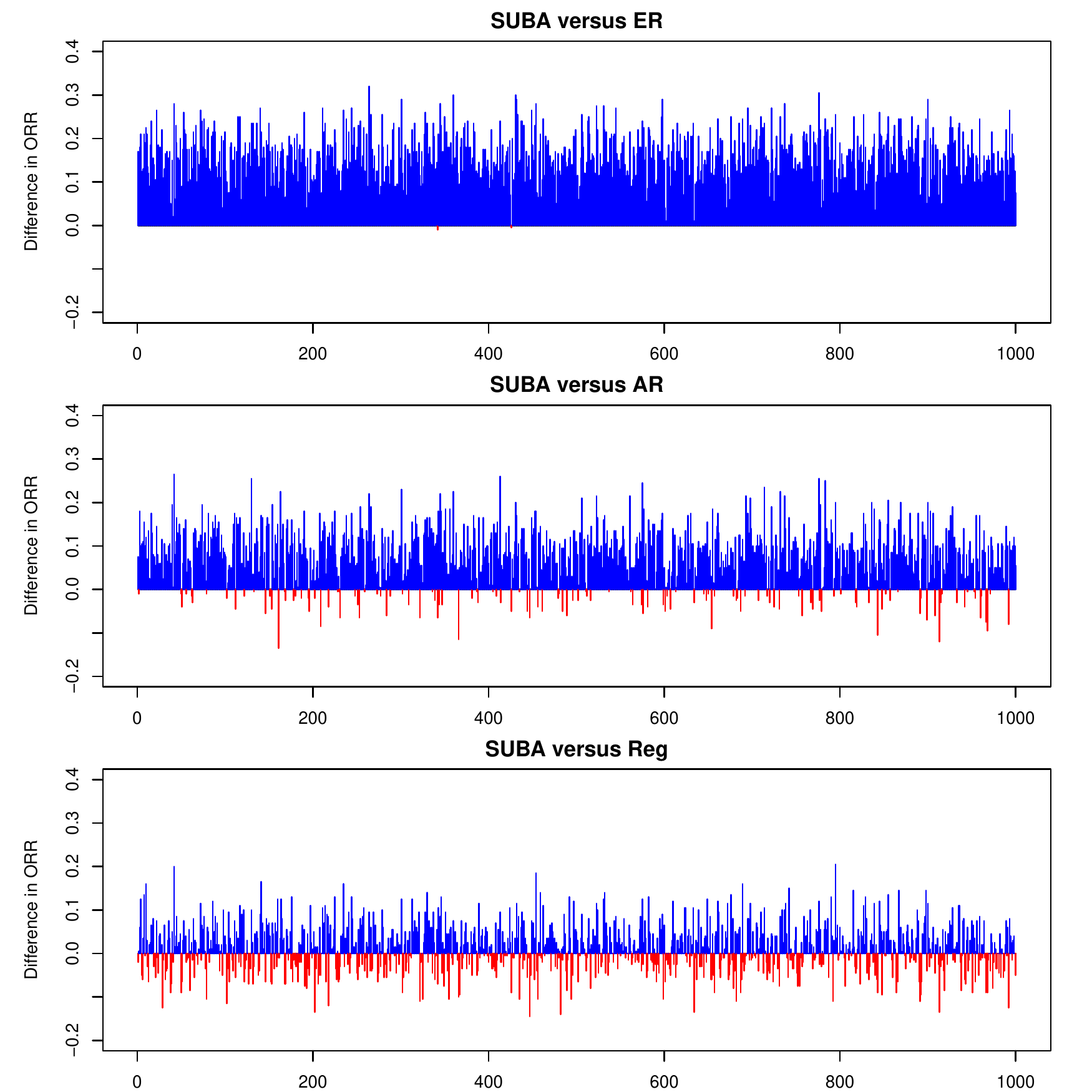}&\includegraphics[scale=0.3]{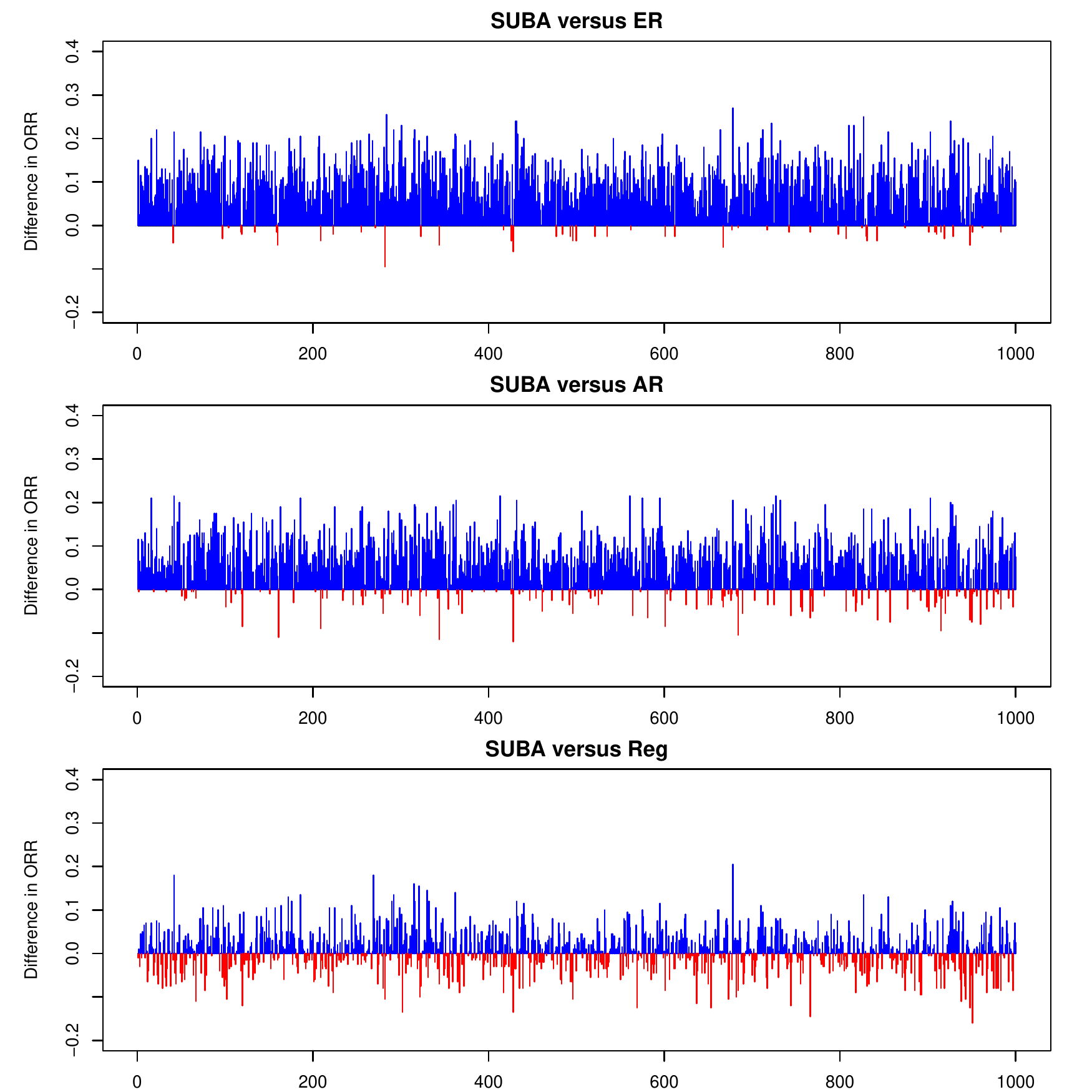}&\includegraphics[scale=0.3]{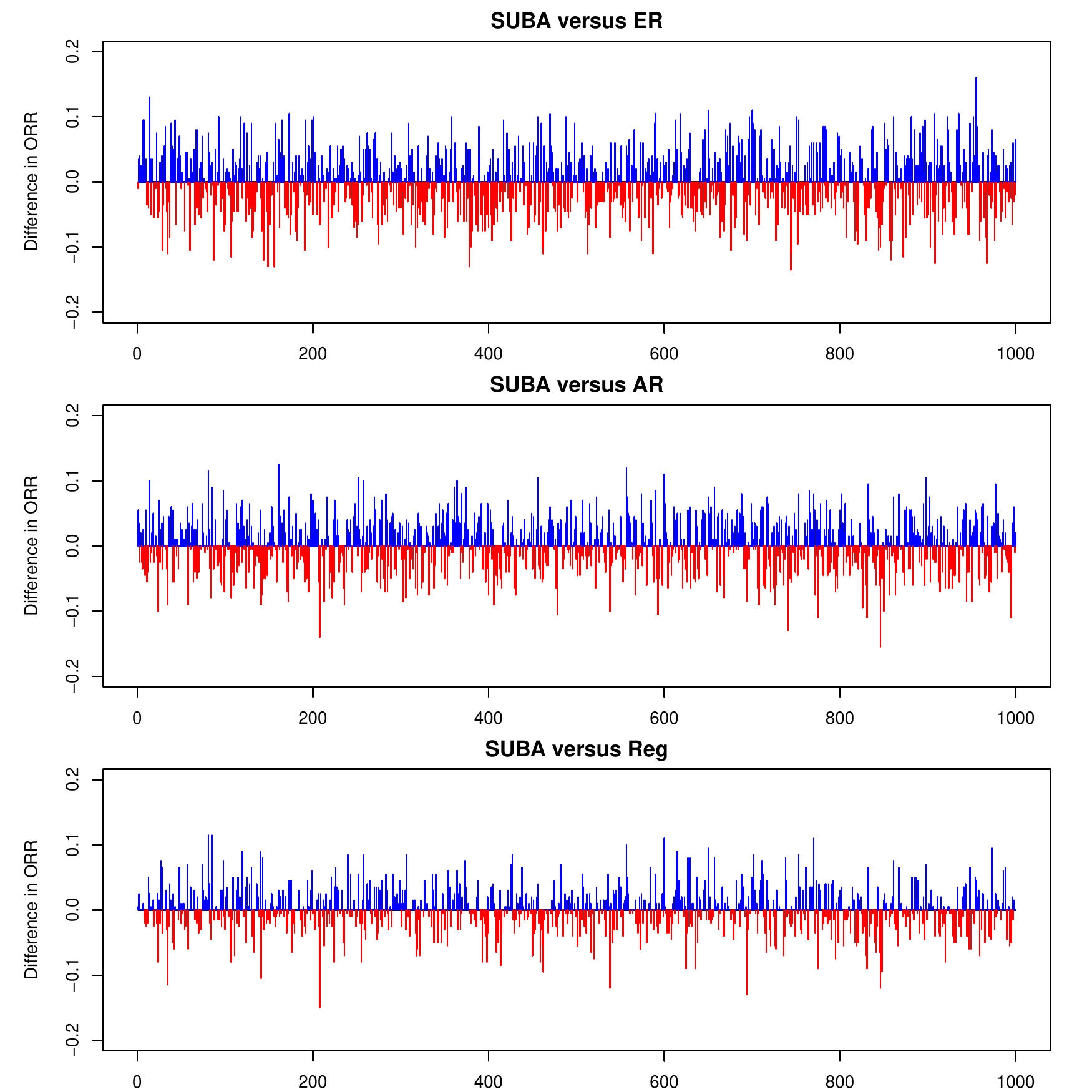}\\
 Scenario 4& Scenario 5& Scenario 6\\
\end{tabular}
\caption{The overall response rate (ORR) comparisons among the ER, AR, Reg and SUBA designs in 1,000 simulated trials in all six scenarios. We plot the ORR differences between SUBA and ER, AR, Reg respectively in each scenario. The blue color represents the ORR of SUBA is higher than ER, AR or Reg; the red color represents lower. }
\label{fig:result2}
\end{figure}

\paragraph*{Response rates.}
Define the overall response rate (ORR) as
$$\mathrm{ORR}=\frac{1}{N-n}\sum_{i=n+1}^NI(y_i=1),$$
which is the proportion of responders among those patients who are treated after the run-in phase. 
We summarize ORR differences between SUBA versus ER, AR, and Reg for each
scenario in Figure \ref{fig:result2}. In  our  comparisons  we     use  the same  burn in period  $n=100$  across  designs.

For \underline{scenarios 2 and 3},  SUBA outperforms ER, AR and Reg with
higher ORR in almost all the simulated trials. The ER and AR designs 
perform similarly.
This suggests  that no gains are obtained  
with AR when the biomarker subgroups are wrongly defined, confirming
      that   for   AR  it  is  essential  an   upfront  appropriate  selection  of  the  biomarker  subgroups. 
 In \underline{scenarios 1, 4 and 5},  SUBA and
Reg are preferable to ER and AR. SUBA exhibits a larger ORR value
than  Reg in 676 of 1,000 simulations in scenario 1, in 612 of 1,000 simulations in scenario 4 and in 605 of 1,000 simulations in scenario 5. 
In \underline{scenario 6}, the true response rates are   constant  and
not related to biomarkers, and   the four designs show similar ORRs
distribution  across 1,000 simulations. 

\paragraph*{Early stopping.}
Table \ref{table1}  reports the average number of patients
under the SUBA design. 
When a trial is stopped early by SUBA, there must be one
last treatment left 
which are considered more efficacious than all the
removed treatments. For a fair comparison with  ER, AR and
Reg which do  not include early stopping, 
summaries in  Table \ref{table2} are based on  assignment  of 
all remaining patients, until the maximum sample size $N$, to that last
active  arm.  

\begin{table}[h]
\centering
\begin{tabular}{ccccccc}
Scenario&1&2&3&4&5&6 \\
\hline
\# of patients&245.28&299.41&300.00&167.63&215.07&209.52\\
\end{tabular}
\caption{The average numbers of patients needed to make the decision of stopping trials early in 1,000 simulated trials in scenarios 1-6. }
\label{table1}
\end{table}

\paragraph*{Treatment assignment.}
We compute the average number of patients (ANP) assigned to treatment
$t$ after the run-in phase by the three designs. Denote
$\mathrm{NP}^d_t$ as the number of patients assigned to treatment $t$
in $d^{th}$ simulated trial after the run-in phase, i.e.,
$\mathrm{NP}^d_t=\sum_{i=n+1}^NI(\zs_i=t)$, $t=1, 2, 3$ and $d=1,
\dots, 1000$. Thus  
$$\mathrm{ANP}_t=\frac{1}{1000}\sum_{d=1}^{1000}\mathrm{NP}^d_t.$$
Table \ref{table2} shows the results. In \underline{scenario 1}, treatment 1 is
always the most effective arm since the second biomarker is fixed at
0.8 (see Figure \ref{fig:simutruth}). We can see  that  most of the patients
are allocated to treatment 1 in scenario 1 by SUBA. 
\underline{Scenario 6} is a
null case in which the biomarkers are not related to response rates
and the response rates   across  treatments are the same, so the
patients allocation by SUBA is similar as ER, AR and Reg.

\begin{table}[h]
\centering
\scalebox{0.8}{
\begin{tabular}{cc|ccc|ccc|ccc|ccc}
\hline
Scenario&&&ER&&&AR&&&Reg&&&SUBA& \\
\hline
&Subset&1&2&3&1&2&3&1&2&3&1&2&3\\
1&/&66.76&66.60&66.64&83.02&65.35&51.63&119.46&70.13&10.41&{\bf 177.11}&18.67&4.22\\
\hline
\multirow{2}{*}{2}&$S_1^0$&33.49&33.09&33.24&33.37&33.19&33.25&35.24&32.88&31.69&{\bf 72.57}&18.37&8.88\\
&$S_2^0$&33.27&33.51&33.40&33.41&33.25&33.53&35.42&33.01&31.76&8.63&17.79&{\bf 73.77}\\
\hline
\multirow{3}{*}{3}&$S_1^0$&19.49&19.09&19.29&22.21&17.63&18.03&18.65&16.40&22.81&{\bf 41.11}&8.94&7.82\\
&$S_2^0$&25.23&25.17&25.35&21.13&26.81&27.80&24.10&21.86&29.79&13.67&{\bf 35.91}&26.17\\
&$S_3^0$&22.05&22.34&22.00&24.61&20.52&21.26&21.27&18.99&26.12&11.33&11.54&{\bf 43.52}\\
\hline
\multirow{2}{*}{4}&$S_1^0$&33.26&33.11&33.44&43.01&42.32&14.49&51.81&48.00&0&{\bf 52.76}&46.96&0.10\\
&$S_2^0$&33.50&33.49&33.20&42.32&43.46&14.41&51.75&48.44&0&50.78&{\bf 49.29}&0.11\\
\hline
\multirow{2}{*}{5}&$S_1^0$&33.26&33.11&33.44&39.14&38.49&22.19&51.51&48.25&0.05&{\bf 51.13}&47.05&1.63\\
&$S_2^0$&33.50&33.49&33.20&38.29&39.32&22.58&51.22&48.92&0.05&47.07&{\bf 51.53}&1.59\\
\hline
6&/&66.76&66.60&66.64&66.66&66.89&66.46&65.04&67.84&67.12&66.90&64.20&68.90\\
\end{tabular}}
\caption{The average numbers of patients (ANPs) assigned to three treatments after the run-in phase in three defined subsets by ER, AR, Reg and SUBA in 1,000 simulated trials in scenarios 1-6. }
\label{table2}
\end{table}

In \underline{scenario 2}, we separately report the average numbers of patients
assigned to three treatments after the run-in phase, among those whose
second biomarker is positive or negative. We separately report these
two averages to demonstrate the benefits of using the SUBA design
since depending on the sign of the second biomarker, different
treatments should be selected as the most beneficial and effective
ones for patients.  When the second biomarker is
positive, treatment 1 is the most superior arm; when the second
biomarker is negative, treatment 3 is the most effective arm according
to our simulation settings.  From Table \ref{table2}, among the
200 post-runin  patients, about 100 patients  have 
($x_{i2}>0$)
 values of the second biomarker. In Table
\ref{table2} we  use $\Sn_1 = \{i:
x_{i2}>0\}$ and $\Sn_2 = \{i: x_{i2}<0\}$ to  denote  sets of patients. 
Think of   $\{\Sn_1, \Sn_2\}$ as a partition in the simulation  truth. 
Among patients in $\Sn_1$,
Table \ref{table2} reports that  an  average  of approximately 73   of  them  are
allocated to treatment 1, 18 to treatment 2, and 9 to treatment 3. For
those in $\Sn_2$, 9 are allocated to treatment 1, 18 to treatment 2, and
74 to treatment 3. Most of the patients are assigned to the
correct superior treatments according to their biomarker values,
highlighting the utility of the SUBA design. In contrast, ER, AR and Reg designs assign far fewer patients to the most effective treatments. 
These  results and,  similarly   Figure \ref{fig:result2},  shows  the  utility  of  the  SUBA  approach.

In \underline{scenario 3}, biomarkers 1, 2 and 3 are related to the response. In
a similar fashion, we report patient allocations by breaking down the
numbers according to three subsets that are indicative of the true
optimal treatment allocation depending on the biomarker
values. Denote $\bar{\theta}_{1i} =
x_{i1}+1.5x_{i2}-0.5x_{i3}+2x_{i1}x_{i3}$,
$\bar{\theta}_{2i}=-x_{i1}-2x_{i3}$, and
$\bar{\theta}_{3i}=x_{i1}-1.5x_{i2}-2x_{i1}x_{i2}$. According to the
simulation truth, we consider three sets $\Sn_1, \Sn_2$ and $\Sn_3$, defined as 
$\Sn_1=\{i: \bar{\theta}_{1i}>\bar{\theta}_{2i} \ \mathrm{and} \
\bar{\theta}_{1i}>\bar{\theta}_{3i}\}$, $\Sn_2=\{i:
\bar{\theta}_{2i}>\bar{\theta}_{1i} \ \mathrm{and} \
\bar{\theta}_{2i}>\bar{\theta}_{3i}\}$ and $\Sn_3=\{i:
\bar{\theta}_{3i}>\bar{\theta}_{1i} \ \mathrm{and} \
\bar{\theta}_{3i}>\bar{\theta}_{2i}\}$. Under this assumption, the
best treatment for patients in set $\Sn_t$ is treatment $t$ according to
the simulation truth. Table \ref{table2} reports the simulation
results for $\Sn_1$, $\Sn_2$ and $\Sn_3$. We can see most of the patients are assigned to the correct superior treatments. In contrast, the ER, AR and Reg designs fail to do so. 

In \underline{scenarios 4 and 5}, biomarkers 1 and 2 are related to the
response. Since treatment 3 is inferior to treatments 1 and 2, the
biomarker space is only split to two sets $\Sn_1$ and $\Sn_2$ according
to simulation truth. Denote $\bar{\theta}_{1i} = x_{i1}^2/2+x_{i1}x_{i2}/2$,
$\bar{\theta}_{2i}=x_{i2}^2/2-x_{i1}x_{i2}/2$. So $\Sn_1=\{i:
\bar{\theta}_{1i}>\bar{\theta}_{2i} \}$ and $\Sn_2=\{i:
\bar{\theta}_{2i}>\bar{\theta}_{1i} \}$. Table \ref{table2} again shows
that 
SUBA assigns more patients to their corresponding optimal treatments 
than ER and AR designs, but performs similar as Reg. Scenarios 4-5
are two challenging cases, in which the dose-response surfaces are
``U''-shaped (plots not shown) and treatments 1 and 2 have similar true responses
rates  for most biomarker values. Treatment 3 is much less desirable
to treatments 1 and 2, and is excluded by SUBA and Reg 
quickly across  most  of the  simulations.  Both designs assign similar numbers of patients on average
to treatments 1 and 2. However, both
designs assign  a considerable number of patients to suboptimal 
treatments. For
example, in both scenarios 50\%
 of the patients received  a  suboptimal treatment, which could be caused
 by false negative splits that failed to capture the superior
 subgroups for those patients. 
 Nevertheless, SUBA is still markedly better than the
ER and AR designs in these scenarios. 

In  summary,  SUBA continuously learns the  response  function  to pair
optimal treatments with targeted patients and   
can  substantially  outperform   ER, AR  and Reg  in  terms of  OOR.  
         
\begin{figure}[h]
\centering
\begin{tabular}{c}
\includegraphics[scale=0.45]{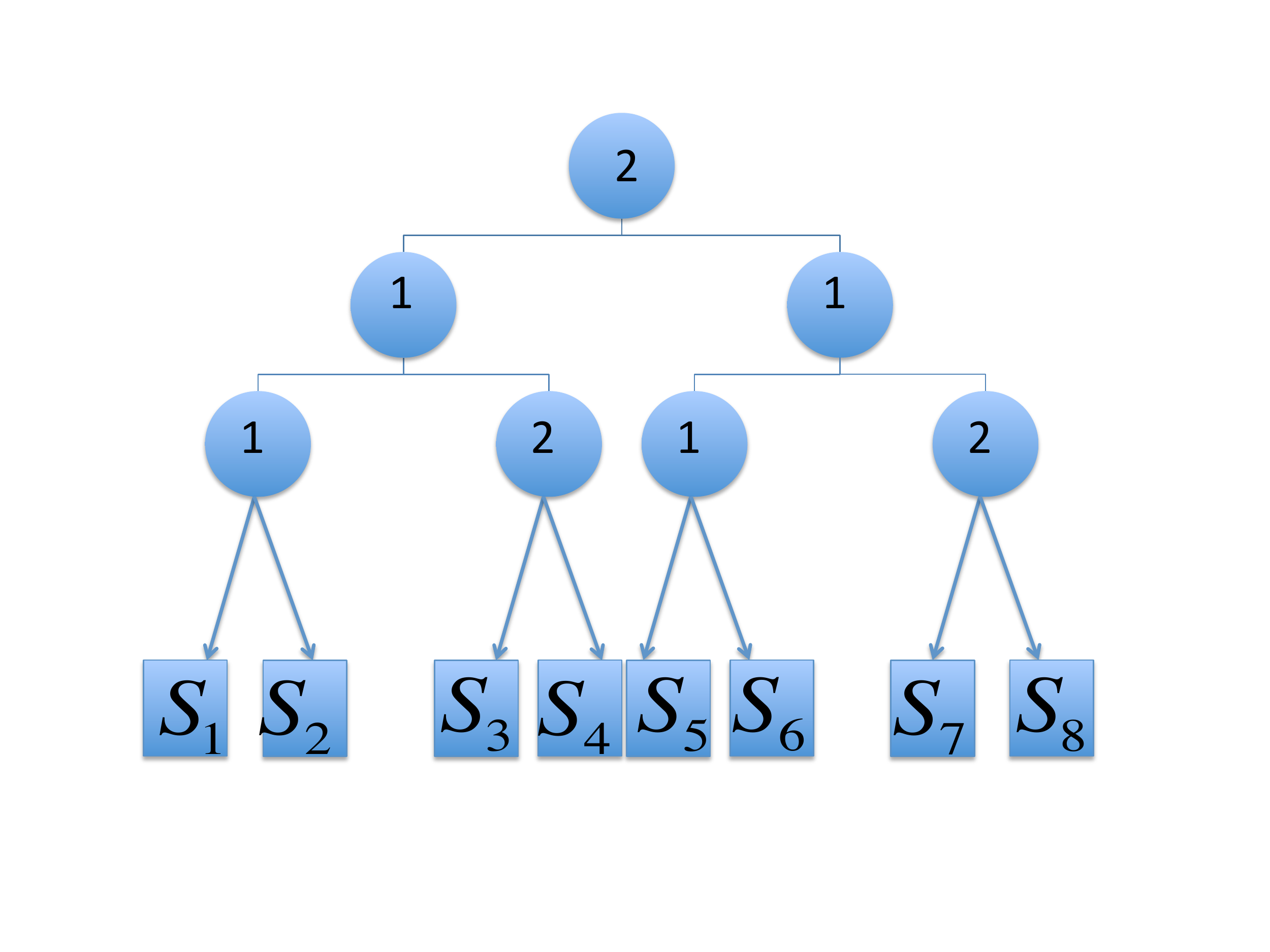}\\
(a) Scenario 2\\
\includegraphics[scale=0.45]{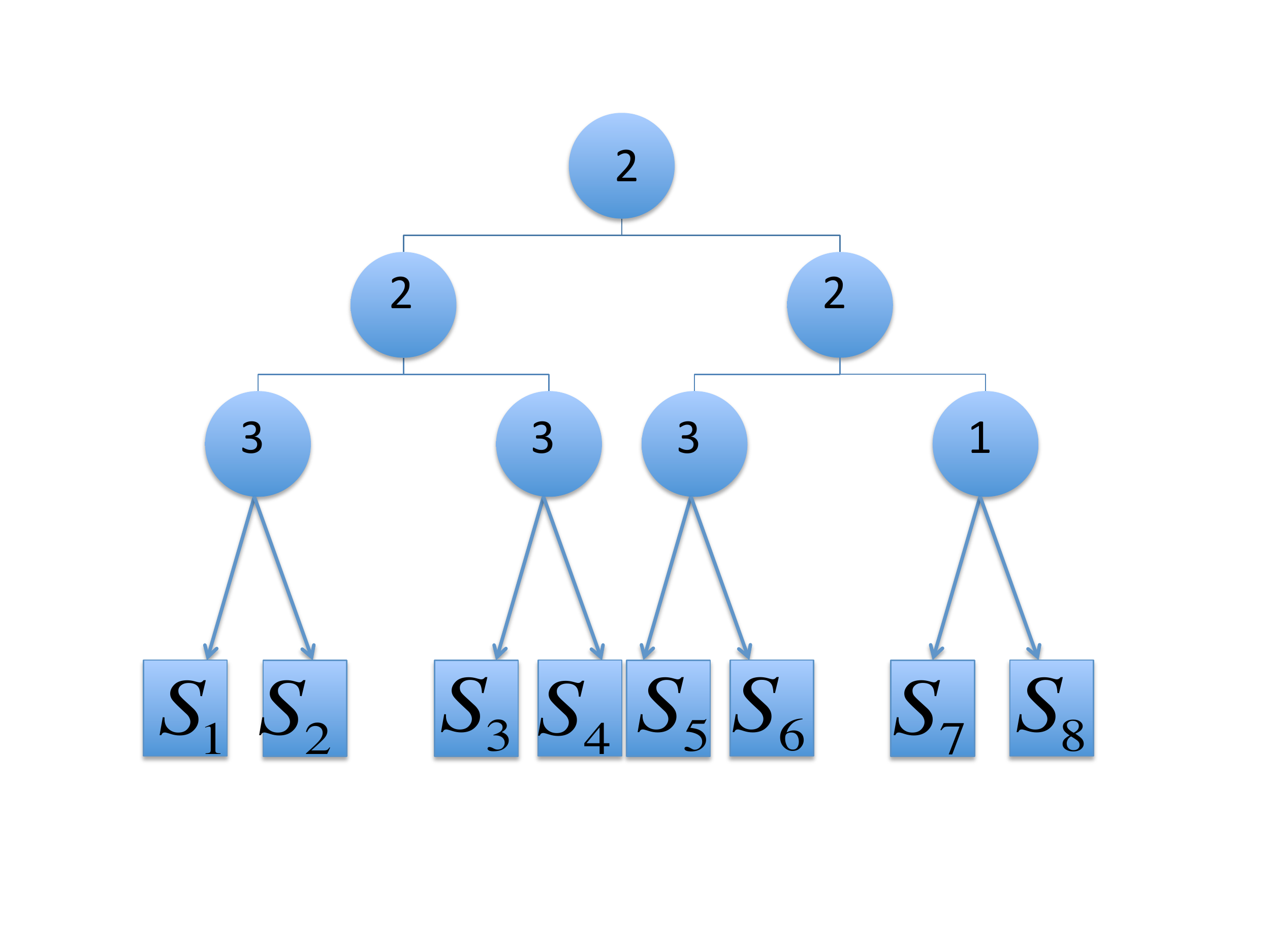}\\
(b) Scenario 3\\
\end{tabular}
\caption{The tree-type least-square partition by SUBA design in one simulated trial in scenarios 2 and 3. The number in the circle represents the biomarker used to split the biomarker space.}
\label{fig:pp}
\end{figure}

\paragraph*{Posterior estimated partition.}
Figure \ref{fig:pp} shows the least-square partition $\Pi^{LS}$ in
an arbitrarily  selected trial for scenarios 2 and 3. The number in
each circle represents the biomarker used to split the biomarker
space. In scenario 2, biomarkers 1 and 2 are related to response
rate. Treatment 1 is the best treatment when the second biomarker is
positive and treatment 3 is the best one when the second biomarker is
negative. The least-square partition $\Pi^{LS}$ uses 
biomarker 2 to split the biomarker space in the first round of split,
which corresponds to the simulation truth. In scenario 3, biomarkers
1, 2, and 3 are related to response rate and the least-square
partition $\Pi^{LS}$ uses these true response-related biomarkers to
split as well. 

\subsection{Sensitivity Analysis}
To evaluate the impact of the maximum sample size on the simulation
results, we carried out a sensitivity analysis with $N=100, 200, 300$ in
scenario 1, with first $n=100$ patients equally randomized. Recall that in scenario 1, treatment 1 has a higher
response rate than treatments 2 and 3, regardless of their biomarker
values. Therefore the effect of sample size on the posterior inference
can be easily evaluated. 

\begin{figure}[Ht!]
 \includegraphics[scale=0.9]{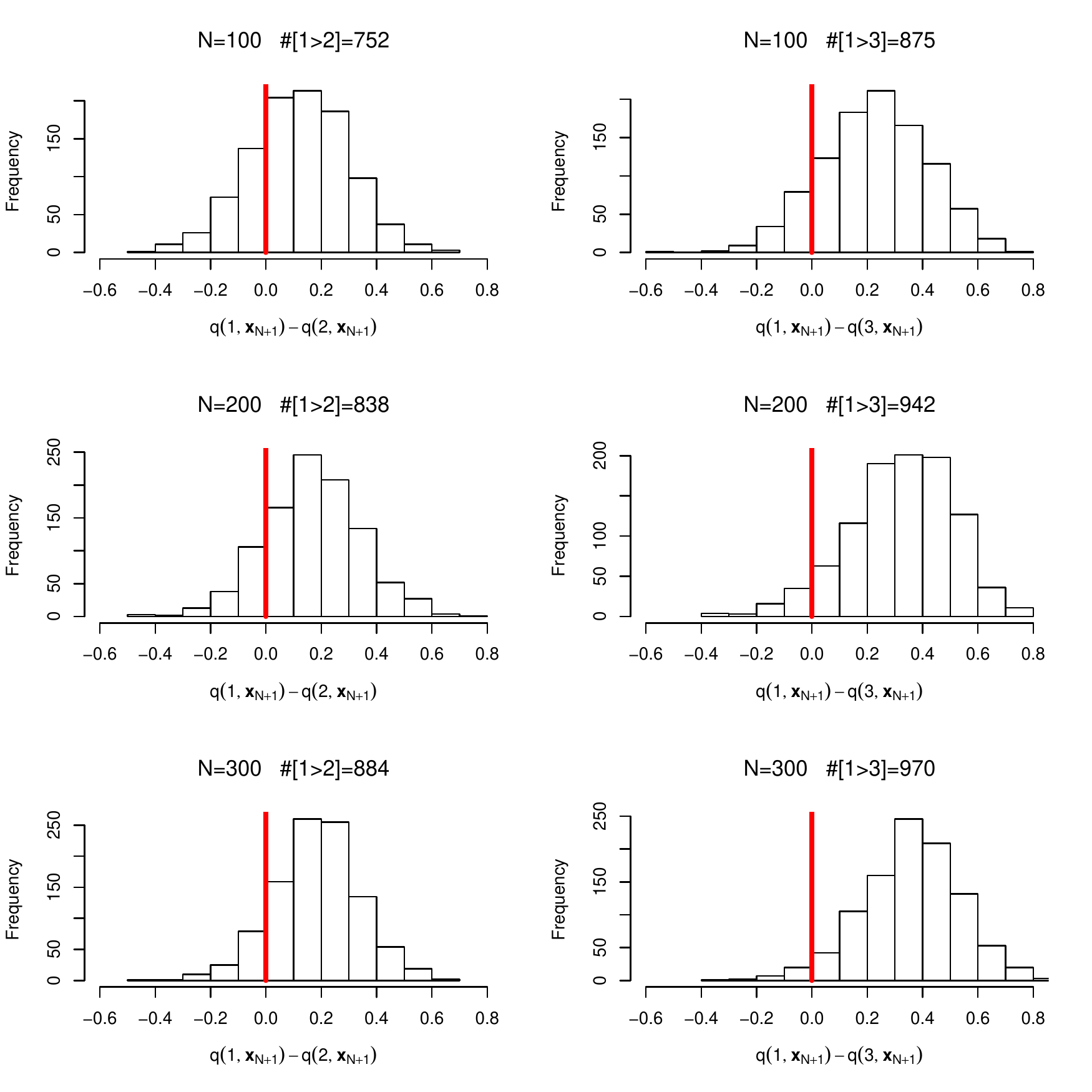}
 \caption{The histogram of $q_{N+1}(1)-q_{N+1}(2)$ and $q_{N+1}(1)-q_{N+1}(3)$ when $N=100, 200, 300$. The right side of red vertical line indicates that the posterior predictive rate of treatment 1 is higher than treatment 2 or treatment 3. }
 \label{fig:sensitivity}
 \end{figure}

Figure \ref{fig:sensitivity} plots the histogram of differences
between treatments $q_{N+1}(1, \xno)-q_{N+1}(2, \xno)$ and
$q_{N+1}(1, \xno)-q_{N+1}(3, \xno)$ after $N=100, 200,$ or $300$
patients have been treated in the trial.  
When $N=100$, treatment 1 is reported as better than treatment 2 in 752 of
1,000 simulations; when $N=200$, treatment 1 is better than treatment
2 in 838 of 1,000 simulations; when $N=300$, treatment 1 is 
better
than treatment 2 in 884 of 1,000 simulations. The more patients
treated, the more precise the posterior estimates and more accurate
assignments for future patients. Similar patterns are observed for the
comparison between treatments 1 and 3. 

We also varied the values $\phi$ and conducted sensitivity
analysis with  $\phi=0.2, 0.5, 0.8$ using scenario 2. Table
\ref{table3} shows the average numbers of patients needed to make the
decision of stopping trials early and the average numbers of patients
assigned to three treatments after the run-in phase in two 
defined subsets.
In summary, the reported summaries vary little across the considered
hyperparameter choices, indicating robustness with respect to changes
within a reasonable range of values.

\begin{table}[h]
\centering
\begin{tabular}{c|ccc|ccc|ccc}
\hline
&&$\phi=0.2$&&&$\phi=0.5$&&&$\phi=0.8$& \\
\hline
\# of patients&&298.10&&&299.41&&&299.15&\\
\hline
Subset&1&2&3&1&2&3&1&2&3\\
$\Sn_1$&{\bf 71.66}&19.09&9.06&{\bf 72.57}&18.37&8.88&{\bf 72.21}&18.50&9.11\\
$\Sn_2$&8.64&18.50&{\bf 73.05}&8.63&17.79&{\bf 73.77}&8.79&18.31&73.09\\
\end{tabular}
\caption{The average numbers of patients needed to make the decision
  of stopping trials early and patient allocation breakdowns 
  in scenario 2 with different values of $\phi=0.2, 0.5, 0.8$. }
\label{table3}
\end{table}


\section{Discussion}
\label{sec:discussion}
We 
  demonstrated \ech the importance of subgroup identification in adaptive
designs when such subgroups are predictive of treatment  responce. The
key contribution of the   proposed model-based approach \ech
is the construction of the random
partition prior $p(\Pi)$ which allows a flexible and simple mechanism
to realize subgroup exploration   as posterior inference on
$\Pi$. \ech
The Bayesian paradigm facilitates   continuous updating \ech of this posterior 
inference   as data becomes available in the trial. \ech
%
The   proposed   construction   for     $p(\Pi)$ \ech   is   easy   to
interpret   and,  most  importantly,   achieve   a  good  balance
between the required computational burden for posterior computation
and the flexibility of the resulting prior distribution. The priors of
$\theta_{t,m}$ are i.i.d Beta$(a,b)$, with $a=b=1$, i.e., a uniform
prior in our simulation studies. If desired, this prior can be
calibrated to reflect the historical response rate of the drug. The
i.i.d assumption simplifies   posterior inference. \ech
Alternatively, one
could impose dependence   across the \ech $\theta$'s; for example, one could
assume that adjacent partition sets have similar $\theta$ values.

The proposed SUBA design focuses on the treatment success for the
patients who are enrolled in the current trial by identifying
subgroups of patients who respond most favorably to each of the
treatments. One could easily add to the SUBA algorithm a final
recommendation of a suitable patient population for a follow-up trial,
such as $\Pi^{LS}$.  
Other directions of generalization \ech
include an extension of the models to incorporate variable selection, when a
large number of biomarkers are measured.  


\section*{Acknowledgment}
The research of YJ and PM is partly supported by NIH R01 CA132897. 
  PM was also partly supported by NIH R01CA157458. 
This research was
supported in part by NIH through resources provided by the Computation
Institute and the Biological Sciences Division of the University of
Chicago and Argonne National Laboratory, under grant S10
RR029030-01. We specifically acknowledge the assistance of Lorenzo
Pesce (U of Chicago) and Yitan Zhu (NorthShore University HealthSystem).

\newpage
\bibliographystyle{chicagoa}
\bibliography{paper}
\clearpage


\end{document}